 \documentclass[authoryear,preprint,review,12pt]{elsarticle}

\usepackage[utf8]{inputenc}
\usepackage[margin=1in]{geometry}
\usepackage{natbib}
\usepackage{enumitem}
\usepackage{amsmath,amssymb}
\usepackage{makecell}
\usepackage{adjustbox}
\usepackage{graphicx}
\usepackage{float}
\usepackage{xcolor}
\usepackage{booktabs}
\usepackage{setspace}
\usepackage{comment}
\usepackage{units}
\usepackage[ruled,vlined]{algorithm2e}
\usepackage[citecolor=blue,colorlinks=true,linkcolor=blue]{hyperref}
\usepackage[hyperref]{ntheorem}
\usepackage{physics}

\DeclareMathOperator*{\argmax}{arg\,max}

\journal{Transportation Research Part B}

\usepackage{xcolor}
\definecolor{MTcommentcol}{RGB}{0,150,0}

\providecommand{\sub}[1]{dummy}
\renewcommand{\sub}[1]{_{\text{\scriptsize{#1}}}}

\begin{document}

\begin{frontmatter}



\title{Vessel-following model for inland waterways based on deep reinforcement learning}

\author[firstAddress]{Fabian Hart\corref{corrAuthor}}
\cortext[corrAuthor]{Corresponding author}
\ead{fabian.hart@tu-dresden.de}
\author[firstAddress]{Ostap Okhrin}
\author[firstAddress]{Martin Treiber}

\address[firstAddress]{Institute of Transportation Economics, Technische Universität Dresden, 01062 Dresden, Germany}

\begin{abstract}
While deep reinforcement learning (RL) has been increasingly applied in designing car-following models in the last years, this study aims at investigating the feasibility of RL-based vehicle-following for complex vehicle dynamics and strong environmental disturbances. As a use case, we developed an inland waterways vessel-following model based on realistic vessel dynamics, which considers environmental influences, such as varying stream velocity and river profile. 
We extracted natural vessel behavior from anonymized AIS data to formulate a reward function that reflects a realistic driving style next to comfortable and safe navigation.  
Aiming at high generalization capabilities, we propose an RL training environment that uses stochastic processes to model leading trajectory and river dynamics. To validate the trained model, we defined different scenarios that have not been seen in training, including realistic vessel-following on the Middle Rhine. Our model demonstrated safe and comfortable driving in all scenarios, proving excellent generalization abilities. Furthermore, traffic oscillations could effectively be dampened by deploying the trained model on a sequence of following vessels. 

\end{abstract}




\end{frontmatter}


\section{Introduction}
\label{sec:introduction}
With the increase of automation in recent decades, autonomous driving technologies have successfully been applied in different traffic domains, e.g., on self-driving cars and unmanned aerial vehicles. Having this knowledge at hand and moving to another transportation domain, current research investigates the feasibility of autonomous vessel traffic on inland waterways, e.g., in \cite{peeters2020unmanned}. A crucial element in the design of autonomous driving technologies, especially in domains with dense traffic, is longitudinal vehicle control, which should guarantee safety and aim for comfortable and economically efficient driving behavior. Various rule-based approaches to model longitudinal vehicle-following have been proposed, e.g., the Gazis-Herman-Rothery model by \cite{gazis1961nonlinear}, or the Intelligent Driver Model (IDM) by \cite{IDM2000Treiber}. The latter has not only been used to model car-following or bicycle-following behavior but was the first step in modeling the longitudinal control of vessels on inland waterways \citep{fischer2014modeling}. 

With the advances in machine learning techniques, various supervised learning approaches have been proposed to model vehicle-following behavior, which relies on the data provided through human demonstrations, e.g., \cite{kuefler2017imitating}, \cite {zhang2011data}. However, since these approaches aim at imitating human driving behavior, this may lead to sub-optimal behavior regarding safety and comfort and will fail at generalization. 

To tackle these problems, current approaches on vehicle-following use deep reinforcement learning (RL), which harnesses the potential of deep neural networks \citep{goodfellow2016deep} and has already shown remarkable achievements in games and real-world problems \citep{mnih2015human, vinyals2019grandmaster, folkers2019controlling,li2021reinforcement}. Instead of imitating human driving behavior, the idea is to optimize predefined safety, efficiency, and comfort metrics directly while interacting with the environment. Some RL-based studies develop training environments where the leading vehicle trajectory is based on real human driver data, such as \cite{HumanLikeAutonomouCF} and \cite{zhu2020safe}.
Similar research proposes a standardized driving cycle serving as a leading vehicle trajectory, used for training, such as \cite{ComparisonRLvsMPC} and \cite{lin2019longitudinal} utilizing the New European Driving Cycle. 
One major drawback coming along with these approaches is that in vehicle-following scenarios, not reflected in the training data set, the performance of the trained model can decrease significantly, revealing inadequate generalization \citep{ComparisonRLvsMPC}. This is also known as the problem of robust out-of-distribution generalization \citep{dittadi2021representation}. To overcome this issue, \cite{hart2021formulation} proposed a stochastic training environment that aims to increase the coverage of possible vehicle-following scenarios, including rare safety-critical situations, such as full-braking of the leader. 

Another issue in the majority of studies investigating RL-based longitudinal vehicle control is the assumption of point-mass kinematic models. There is only a handful of approaches that consider complex vehicle dynamics, such as {\cite{lin2019longitudinal} and \cite{farag2020reinforcement}}. Furthermore, existing research does not consider environmental disturbances that can impact vehicle dynamics. This motivated us to study the feasibility of RL-based vehicle-following models for complex vehicle dynamics under major environmental influences. To investigate this issue, we chose vessel-following on inland waterways as our application domain where environmental disturbances, such as varying water stream dynamics and changing river geometry, strongly impact the vessel dynamics.
To the best of our knowledge, there exist no vessel-following models for inland waterways based on reinforcement learning. Most of the work in this area focuses on convoys or platoons in open water \citep{zhang2019multi, liang2021platoon}.

The desired behavior of the trained RL model depends on the reward function, which has to be designed beforehand. Typically, these reward functions are hand-crafted and rely on the expertise of designers, who, for example, define safe distances or headways to the leader vehicle, cf. \cite{DampenStopAndGoTraffic}, \cite{ CFelectricVehicle}, \cite{ masmoudi2021reinforcement}. Since, in contrast to road traffic, commonly used values for safe distances or headways to the leader vehicle do not exist in vessel traffic, we are adapting the solution of \cite{zhu2020safe} by evaluating human driving data that we extract from the Automatic Identification System (AIS) system. 

Summarizing, the main contribution of our work is to investigate the feasibility of RL-based vehicle-following models for complex vehicle dynamics and under strong environmental influences, which includes the following aspects for the first time considered jointly:
\begin{itemize}
    \item Extraction of realistic vessel behavior from anonymized AIS data.
    \item Formulation of a reward function, using extracted behavior and reflecting safety and comfort aspects.
    \item Design of an RL training environment, considering realistic vessel dynamics and environmental disturbances, with a focus on high generalization capabilities to tackle the problem of out-of-distribution robustness.
    \item Training of an RL vessel-following model and scenario-based validation regarding safety, efficiency, and comfort. 
    \item Evaluation of the generalization capabilities by simulating realistic scenarios on the Middle Rhine and testing for string stability.
\end{itemize}

This work is structured as follows: In Section \ref{sec:RLmethodology}, we introduce the RL methodology. In Section \ref{sec:approach}, we propose our approach for a RL-based vessel-following model using realistic vessel dynamics under environmental disturbances. We validate the trained model in Section \ref{sec:validation}, followed by a conclusion in Section \ref{sec:Conclusion}.

\section{Reinforcement learning methodology}
\label{sec:RLmethodology}
\subsection{Basics}\label{subsec:RL_basics}

The objective of RL is to solve sequential decision tasks where an agent interacts with the environment, maximizing the discounted cumulative reward \citep{sutton2018reinforcement}. The formal basis of RL are Markov decision processes that consists of a state space $\mathcal{S}$, an action space $\mathcal{A}$, an initial state distribution $T_0: \mathcal{S} \rightarrow [0,1]$, a state transition probability distribution $\mathcal{P}: \mathcal{S} \times \mathcal{A} \times \mathcal{S} \rightarrow [0,1]$, a reward function $\mathcal{R}: \mathcal{S} \times \mathcal{A} \rightarrow \mathbb{R}$, and a discount factor $\gamma \in [0,1]$. After receiving a state information $s_t \in \mathcal{S}$ at each time step $t$, the agent selects an action $a_t \in \mathcal{A}$, gets an instantaneous reward $r_{t+1}$, and transitions based on the environmental dynamics $\mathcal{P}$ to the next state $s_{t+1} \in \mathcal{S}$. In the following, we use capital notation, e.g., $S_t$, to indicate random variables and small notation, e.g., $s_t$ or $s$, to describe their realizations.  

The RL agent aims to learn a policy $\pi$, that is a mapping from each state $s \in \mathcal{S}$ to an action $ a \in 
 \mathcal{A}$ in order to maximize the expected discounted cumulative reward, starting from state $S_0$: $J(\pi) =  E_{\pi}\left[ \sum_{k=0}^{\infty} \gamma^k R_{k+1}\right | S_0 = s]$. The definition of action value functions $Q^{\pi}(s,a)$ as the expected return when starting in state $s$, taking action $a$, and following policy $\pi$ afterwards, is a key function: $Q^{\pi}(s,a) = E_{\pi}\left[ \sum_{k=0}^{\infty} \gamma^k R_{t+k+1} | S_t = s, A_t = a \right]$. Furthermore,  $\pi^*(s) = \argmax_{a \in \mathcal{A}} Q^*(s,a)$ is defined as a deterministic optimal policy, that is linked with an optimal action-value function $Q^*(s,a) = \max_{\pi} Q^{\pi}(s,a)$. To learn $Q^*(s,a)$, the use of the \cite{bellman1954theory} optimality equation is common practice: 
\begin{equation}\label{eq:Bellman_opt_eq}
    Q^*(s,a) = \mathcal{R}(s,a) + \gamma \sum_{s' \in \mathcal{S}} \mathcal{P}_{sa}^{s'} \max_{a' \in \mathcal{A}} Q^*(s',a').
\end{equation}
Based on the Bellman equation, \cite{watkins1992q} introduced the popular $Q$-learning algorithm, where $Q$-values are approximated by tabular representations. This allows storing a finite number of $(s,a)$-pairs, which restricts the algorithm to discrete state spaces. To allow for continuous state spaces, the $Q$-values are approximated by more complex representations like deep neural networks. Based on this approach, \cite{mnih2015human} introduced the deep $Q$-network (DQN) algorithm, combining $Q$-learning with function approximation. The training of the function $Q^{\omega}(s,a)$ with parameter vector $\omega$ is realized by gradient descend algorithm:
\begin{equation}\label{eq:DQN_grad_descent}
    \omega \leftarrow \omega + \alpha \left\{y-Q^{\omega}(s,a) \right\} \nabla_{\omega} Q^{\omega}(s,a), 
\end{equation}
with reward $r$, target $y = r + \gamma \max_{a' \in \mathcal{A}} Q^{\omega'}(s',a')$, and learning rate $\alpha$. $Q^{\omega'}(s,a)$, named as the target network, defines a time-delayed copy of the original network with parameter $\omega$. This technique has been found to enhance the stability of the training process.
Another feature of DQN is experience replay, which is used to sample transitions randomly (or with more advanced strategies like \cite{schaul2015prioritized}) to perform gradient descent steps. 
Since learning with DQN is based on calculating the maximum over all possible actions, this algorithm just allows for discrete action spaces $\mathcal{A}$. As our application case involves continuous action spaces, we use the deep deterministic policy gradient (DDPG) algorithm \citep{lillicrap2015continuous}.

\subsection{Deep deterministic policy gradient (DDPG)}
The DDPG algorithm has been proven to perform well on control problems with continuous state and action spaces that are similar to our task, such as \cite{HumanLikeAutonomouCF}, \cite{lin2019longitudinal}, and \cite{du2021optimized}.
 DDPG is an off-policy, actor-critic algorithm based on an actor function $\mu^{\theta}: \mathcal{S} \rightarrow \mathcal{A}$ with parameter vector $\theta$ that approximates the maximum operation in the target computation, and the critic function $Q^{\omega}(s,a)$ that approximates the action-values like in the DQN algorithm. In this context, the critic evaluates the actions made by the actor. Both actor and critic functions are represented as neural networks.
 
  We consider the performance objective $J(\mu^{\theta}) = E_{\mu^{\theta}}\left[ \sum_{k=0}^{\infty} \gamma^k R_{k+1}\right | S_0]$ based on the deterministic policy $\mu^{\theta}$. \cite{silver2014deterministic} proved the \emph{Deterministic Policy Gradient Theorem}, which yields the gradient of the performance measure with respect to $\theta$:
\begin{equation}\label{eq:deter_pol_grad_theorem}
    \nabla_{\theta}J(\mu^{\theta}) \approx E_{s \sim \rho^{\mu}} \left\{\nabla_{\theta} \mu^{\theta}(s) \nabla_a Q^{\omega}(s,a)|_{a=\mu^{\theta}(s)} \right\},
\end{equation}
where $\rho^{\mu}$ is the discounted state visitation distribution. In the DDPG algorithm, this gradient is used to train the actor via gradient ascent. We refer to \cite{lillicrap2015continuous} for more details.
 Furthermore, experience replay and target networks from DQN are adapted with a minor adjustment. By applying a soft-update of the target networks for both actor and critic, the update targets change slowly, which has been found to enhance training stability. Denoting $\tau$ as the soft target update rate, $\theta'$ and $\omega'$ the parameter sets of the target actor and critic, respectively, the update is:
\begin{align}\label{eq:DDPG_soft_tgt_up}
    \omega' &= \tau \omega + (1-\tau) \omega', \nonumber \\ 
    \theta' &= \tau \theta + (1-\tau) \theta'.
\end{align}
The complete algorithm is detailed in Algorithm \ref{algo:DDPG}.\\

\begin{algorithm}[H]
\setstretch{1.05}
\SetAlgoLined
 Randomly initialize actor $\mu^{\theta}$ and critic $Q^{\omega}$ \\
 Initialize target actor $\mu^{\theta'}$ and target critic $Q^{\omega'}$ with  $\theta^{'} \leftarrow \theta$ and $\omega' \leftarrow \omega$  \\
 Initialize replay buffer $\mathcal{D}$\\
  \For{episode = 1,M}{
  Initialize a random process $\mathcal{N}$ for action exploration\\
 Receive initial state $s_0$ from environment \\
 \For{t = 1,T}{
 \emph{Acting}\\
 Select action $a_t = \mu^{\theta}(s_t) + \mathcal{N}_t$  according to current policy \\
 Execute $a_t$, receive reward $r_{t+1}$, new state $s_{t+1}$, and done flag $d_t$\\
 Store transition $(s_t, a_t, r_{t+1}, s_{t+1}, d_t)$ to $\mathcal{D}$\\
 \medskip
 \emph{Learning}\\
  Sample random mini-batch of transitions $(s_i, a_i, r_{i+1}, s_{i+1}, d_i)_{i=1}^{N} $ from $\mathcal{D}$\\
  Calculate target:\\
  \vspace{-1cm}
  \begin{align*}
      y_i &= r_{i+1} + \gamma (1-d_i) Q^{\omega'} \left\{ s_{i+1},\mu^{\theta'}(s_{i+1})\right\}
  \end{align*} \\
 Update critic by minimizing loss: $L =  N^{-1} \sum_{i=1}^{N} \left\{y_i - Q^{\omega}(s_i, a_i)\right\}^2$\\
   \vspace{0.2cm}
 Update actor policy using the sampled policy gradient:\\ 
   \vspace{-0.7cm}
  \begin{align*}
  \nabla_{\theta}J \approx \sum_{i=1}^{N} \nabla_a Q^{\omega}(s,a)|_{s=s_i, a=\mu^{\theta}(s_i)} \nabla_{\theta} \mu^{\theta}(s)|_{s=s_i}
  \end{align*} \\
  Update target networks via (\ref{eq:DDPG_soft_tgt_up})
 
 \medskip
\emph{End of episode handling}\\
    \If{$d_t$}{
    Reset environment to an initial state $s_{t+1}$\\
    }
    }
}
 \caption{DDPG algorithm following \cite{lillicrap2015continuous}}
 \label{algo:DDPG}
\end{algorithm}

\subsection{Architecture and Hyperparameters}
\label{subsec:implementation}
Both neural networks, representing actor $\mu^{\theta}$ and critic function $Q^{\omega}(s,a)$, are feed-forward neural networks with two layers of hidden neurons, containing 32 neurons each. ReLU activation functions (\cite{relu}) are used for all layers, except for the output layer of the actor network that uses a $tanh(\cdot)$ activation function. The learning rates for updating the weights of the critic and actor network, $\alpha_{\rm actor}$ and $\alpha_{\rm critic}$, are set to 0.001. Optimization is performed with Adam \citep{kingma2014adam}. As suggested in \cite{lillicrap2015continuous}, we used temporally correlated noise to explore well in physical environments with momentum. We adapted an \cite{OU} process with $\theta_{OU} = 0.15$ and $\sigma_{OU} = 0.2$. The complete list of hyperparameters is given in Table \ref{tab:hyperparams}.
\begin{table}[H]
    \centering
    \begin{tabular}{l|l}
    Hyperparameter & Value\\
    \toprule
        Discount factor $\gamma$ &  0.95 \\
        Batch size $N$ & 32\\
        Replay buffer size $|\mathcal{D}|$ & $10^5$ \\
        Learning rate actor $\alpha_{\rm actor}$ & $0.001$ \\ 
        Learning rate critic $\alpha_{\rm critic}$ & $0.001$ \\ 
        Soft target update rate $\tau$ & 0.001 \\
        Optimizer & Adam \\
        Exploration noise $\theta_{OU}$ & 0.15 \\
        Exploration noise $\sigma_{OU}$ & 0.2 \\
       	Number of hidden layers & 2 \\
		Neurons per hidden layer & 32 \\
    \end{tabular}
    \caption{List of DDPG hyperparameters. }
    \label{tab:hyperparams}
\end{table}

\section{Approach: Vessel-Following}
\label{sec:approach}
\subsection{Problem Description}
\label{sec:problemDescription}
Vehicle-following refers to safe, efficient, and comfortable following of a leader vehicle. As motivated in Section \ref{sec:introduction}, we aim to investigate the feasibility of an RL-based vehicle-following model for complex vessel dynamics and environmental influences, such as river flow dynamics with changing river geometry. Thereby, the following vessel is controlled by an RL agent that sets a suitable value for the engine power for each time step. This engine power is then translated into an acceleration under consideration of the vessel dynamics and river influences. 

\subsection{Vessel Dynamics}
\label{sec:vesselDynamics}
The mathematical vessel model, used in this work, is based on \cite{linke2015farao} who defined the basic longitudinal equation of vessel motion as the Newtonian momentum balance:
\begin{equation}
\label{eq:momentumBalance}
\frac{{\rm d}}{{\rm d}t}  \left(m_{\rm eff} \dot{x}\right)=m_{\rm eff}\ddot{x}+\dot{x}\dot{m}_{\rm eff}=T_{prop}-W\sub{hyd}-W\sub{hull}-W_g.
\end{equation}
The rate of change of the momentum with the dynamic mass $m_{\rm eff}$ and the vessel speed $\dot{x}$ is equal to the thrust $T_{prop}$ by the propellers subtracted by the drag resistance $W_{\rm hyd}$ of moving through the water, the resistance $W_{\rm hull}$ from friction between the water and the hull of the vessel, and the momentum transfer $W_g$ of the river and induced currents which also includes the gravitational pull by the slope of the water surface. The thrust $T$ depends on the engine power $P$ and the speed 
$v_r=\dot{x}-v_{\rm str}$ relative to the water stream speed $v_{\rm str}$. The resistances depend on $v_r$, $v_{\rm str}$, the draft (vertical distance between the waterline and the bottom of the hull), and the river geometry, which we approximate by a rectangular profile of
depth $h$ and width $w$ resulting in a cross-section $A_{cross}=w h$, for more details we refer to \cite{linke2015farao}. Note that we do not consider lateral dynamics in this work. Furthermore, we use a typical inland cargo vessel type with the mass  $m = \unit[3174]{t}$, the length $L_s = \unit[110]{m}$, the width $W_s=\unit[11.4]{m}$ and the draft $H_s=\unit[2.8]{m}$.

\subsection{Action, State and Reward}
\label{sec:actionStateReward}
At each time step $t$, the agent computes a continuous action $a_t \in [0,1]$ that is mapped to an engine power $P_t$:
\begin{equation}
    P_t = P_{\rm max} a_t,
\end{equation}
where $P_{\rm max}$ defines the maximum possible engine power. Note that we do not consider negative engine powers in this study. To make adequate decisions, the agent must observe its environment. From a sensory point of view, we assume that the agent is able to perceive the bow-to-stern gap $g_t$ and relative speed $ \Dot x_t - \Dot x_{t,lead}$ to the leader vessel  at time step $t$. Furthermore, the agent senses the current river depth below keel $h_{t}$, the cross-sectional river area $A_{t,cross}$ and water stream speed $v_{t,str}$. Summarizing, the agent observes the state $s_t$ at time step $t$ that is defined as:
\begin{equation}
\begingroup
    \renewcommand*{\arraystretch}{1.2}
        s_t = \left(
        \frac{\Dot x_t }{v_{\rm scale}},
        \frac{P_t}{P_{\rm max}},
        \frac{g_t}{g_{\rm scale}},
        \frac{\Dot x_t - \Dot x_{t,lead}}{v_{\rm scale}},
        \frac{h_{t}}{h_{\rm scale}},
        \frac{A_{t,\rm cross}}{A_{\rm scale}},
        \frac{v_{t,\rm str}}{v_{\rm scale}}\right),
    \endgroup
\end{equation}
where the parameters $v_{\rm scale}$, $P_{\rm max}$, $g_{\rm scale}$,  $h_{\rm scale}$ and  $A_{\rm scale}$ are used for normalizing the observations, with the values to be found in Table \ref{tab:EnvsParameters}.

As motivated in Section \ref{sec:introduction}, we use real human driving data from the Automatic Identification System (AIS) to extract driving behavior. As the AIS system is obligatory for many inland waterways, AIS data is used in various RL-based applications to extract vessel trajectories, such as in \cite{guo2020autonomous} and \cite{westerlund2021learning}. In this study, we aim to extract vessel-following behavior from the AIS database. We chose a section from the Middle Rhine in Germany for two purposes: First, this part of the Rhine is quite narrow, so overtaking maneuvers are relatively rare. Second, the traffic volume is high. These characteristics lead to a higher chance of vessel-following events that we try to extract. 

\begin{figure}[]
    \centering
    \includegraphics[width=1\linewidth]{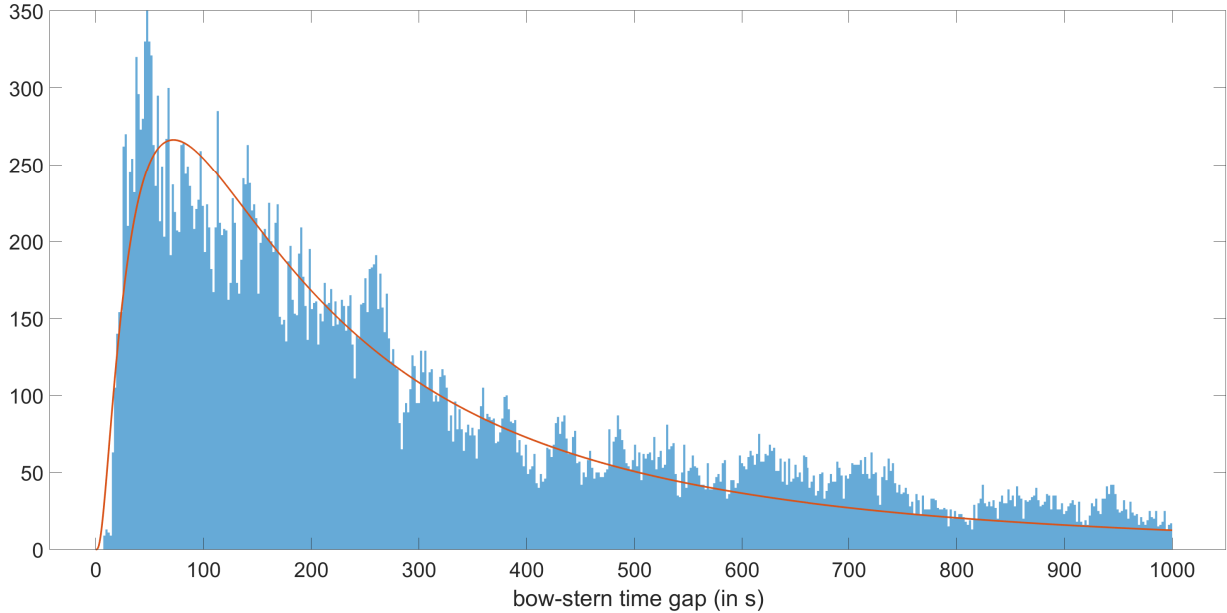}
    \caption{Distribution of bow-stern time gap $T$ in extracted vessel-following events from the Middle Rhine for a window of \unit[60]{km} and \unit[24]{h}. A lognormal distribution was fit on the data to model the reward function.}
    \label{fig:timegap_histo}
\end{figure}

We defined a section of $\unit[60]{km}$ length and a time span of $\unit[24]{hours}$ as an observation window. Within this window, we extracted vessel-following events, characterized by two criteria: First, the difference speed between follower and leader vessel must be below a threshold of $\unit[0.2]{m/s}$. Second, the follower and leader vessel must have a lateral overlap. Based on the extracted vessel-following events, we adapted the approach of  \cite{zhu2020safe} by analyzing the follower's bow-stern time gap $T$ and fitting a distribution onto it to model a part of the reward function. For computing $T$, we used the vessel's speed with respect to the current water stream speed. A histogram of  the bow-stern time gap  $T$ of all extracted vessel-following events in the interval $T \in [0, 1000]$ is depicted in Figure \ref{fig:timegap_histo}.
It can be noted that the resulting distribution of the bow-stern time gap for vessels looks relatively similar to the distribution in road traffic \citep{wasielewski1979car}. To approximate the data, a lognormal distribution was fit onto it, with its density function defined as:
\begin{equation}
f_{\text {T }}(T \mid \mu_{\rm T}, \sigma_{\rm T})=\frac{1}{T \sigma_{\rm T} \sqrt{2 \pi}} \exp\left\{\frac{-(\ln T-\mu_{\rm T})^{2}}{2 \sigma_{\rm T}^{2}}\right\},
\end{equation}
where $\mu_{\rm T}$ and $\sigma_{\rm T}$ are the mean and log standard deviation of the lognormal distribution. The resulting values for these parameters are given in Table \ref{tab:EnvsParameters}.

Adopted from \cite{zhu2020safe}, we model the safety factor of the reward function based on the approximated distribution and the bow-stern time gap $T_t$ at time step $t$ as:
\begin{equation}
\label{eq:timeheadwayReward}
    r_{t,\rm safety} =  f_{\rm T}(T_t \mid \mu_{\rm T}, \sigma_{\rm T}).
\end{equation}
Next to keeping a safe headway, this reward factor also aims to improve traffic flow efficiency by motivating the follower vessel to keep not too long headways. 

Apart from safety and efficiency, vehicle-following models consider comfortable driving. In this study, we characterize comfort as low changing rates in engine power $P$ which result in low accelerations. In this sense, we define the comfort factor of the reward function at time step $t$ as: 
\begin{equation}
    r_{t,\rm comfort} =- \left(\frac{\Delta t}{P_{\rm max}} \dv{P_t}{t}\right)^2.
\end{equation}

Summarizing, the final reward at simulation time step $t$ is defined as the weighted sum of the two reward factors according to:
\begin{equation}
    r_t =  r_{t,\rm safety} + \beta r_{t,\rm comfort},
\end{equation}
where the weight $\beta$ has been found experimentally. Its value can be  found in Table \ref{tab:EnvsParameters}. 

\subsection{Training Environment}
\label{sec:trainingEnv}
As outlined in the introduction, the objective of this study is to train an agent that is capable of handling vessel-following scenarios in different environments, in particular, for different river characteristics and leading trajectories. Instead of training the agent on the Middle Rhine and in scenarios based on AIS data, we aim for a more generic training environment. To achieve this, we are adapting the approach of \cite{hart2021formulation, hart2021impact} by using general stochastic processes to design environmental influences. This method proved to yield good generalization capabilities. In detail, we use AR(1) processes \citep{tsay2005analysis} to model leading vessel trajectory and river characteristics in training:
\begin{align} 
	X_{t+1}&=c_{\rm AR}+ \phi_{\rm AR} X_t + u_t, &\text{where} \quad u_t \sim \mathcal{N}(0, \sigma_{\rm AR}^2),
\end{align}
with auto-regressive parameters $c_{\rm AR}$ and $\phi_{\rm AR}$ and variance $\sigma_{\rm AR}^2$. 

For each training episode, we define an independent AR(1) process for leading vessel speed $\Dot x_{t,\rm lead}$ and river observations as river depth below keel $h_{t}$, cross-sectional river area $A_{t,\rm cross}$, and water stream speed $v_{t,\rm str}$ at time step $t$ as:
\begin{align}	
	\Dot x_{t+1,\rm lead} &= c_1 + \phi_{1} \Dot x_{t,\rm lead} + u_{t,1},  &\text{where} \quad u_{t,1} \sim \mathcal{N}(0, \sigma_{1}^2),\\
	h_{t+1} &=c_2 + \phi_{2} h_{t} + u_{t,2}, &\text{where} \quad u_{t,2} \sim \mathcal{N}(0, \sigma_{2}^2),\\
	A_{t+1,\rm cross} &=c_3 + \phi_{3} A_{t,\rm cross} + u_{t,3}, &\text{where} \quad u_{t,3} \sim \mathcal{N}(0, \sigma_{3}^2),\\
	v_{t+1,\rm str} &=c_4 + \phi_{4} v_{t,\rm str} + u_{t,4}, &\text{where} \quad u_{t,4} \sim \mathcal{N}(0, \sigma_{4}^2),
\end{align}
with $\phi_i$ and $\sigma_i^2$ for $i = 1,\dots,4$ defining auto-regressive parameters and variances. These parameters have been adjusted to cover reasonable ranges and changing rates of the respective variable. Their values can be found in Table \ref{tab:EnvsParameters}. Since a vessel, having relative speeds with respect to water stream lower than $\unit[2]{m/s}$, is not maneuverable, we constrain $\Dot x_{t,lead}$ to that lower bound by setting the relative speed to values $\ge \unit[2]{m/s}$ within the process. In the same way we constrain the river depth below keel $h_{t}$ to be $\ge 0$.
Notice that, in spite of ignoring the lateral dynamics, the river cross-section enters via the back-current terms of the resistance forces in Eq.~\eqref{eq:momentumBalance}.

To simulate an episode, the vessel dynamics (\ref{eq:momentumBalance}) have to be integrated. One training episode covers 500 time steps, and the Euler and ballistic methods are used to update the speed and position for time step $t + 1$, respectively. This approach is recommended in \cite{numericalUpdateMethodsTreiber} as an efficient and robust scheme for integrating car-following models:
\begin{align}
	\dot x_{t+1} &= \dot x_{t} + \ddot x_{t} \Delta t, \\
	x_{t+1} &= x_{t} + \frac{\dot x_{t} + \dot x_{t+1}}{2} \Delta t,
\end{align}
with $\Delta t$ corresponding to the simulation step. To initialize an episode, we set $P_0$ to zero and $\dot x_0$ and $\dot x_{0,lead}$ is sampled uniformly from the interval $[\unit[2]{m/s}, \unit[6]{m/s}]$. Since we aim to train an agent that is applicable for free-driving, approaching, and vessel-following scenarios, we further set the initial gap $g_0$ between both vessels to $\unit[600]{m}$, so that approaching of the leading vessel is part of an episode.

\begin{table}
\caption{Description and value for environment parameters}
\label{tab:EnvsParameters} 
\begin{center}
	\begin{tabular}{ p{0.1\linewidth} p{0.6\linewidth} p{0.15\linewidth} } 
		Parameter & Description & Value   \\ \hline
		$P_{\rm max}$ & maximum  possible  engine  power  & $\unit[1]{MW}$\\
		$v_{\rm scale}$ & speed scaling parameter & $\unit[6]{m/s}$\\
		$g_{\rm scale}$ & gap scaling parameter & $\unit[800]{m}$\\
		$h_{\rm scale}$ & river depth scaling parameter & $\unit[3]{m}$\\
		$A_{\rm scale}$ & cross-sectional area scaling parameter & $\unit[1500]{m}$\\
		$\Delta t$ & simulation step size & $\unit[1]{s}$\\
		$\mu_{\rm T}$ 		& mean of lognormal distribution  			& $\unit[5.41]{}$  \\
		$\sigma_{\rm T}$ 		& log standard deviation  	    	& $\unit[1.06]{}$   \\
		$\beta$       & weighting factor in reward function   & $0.0004$ \\
	    $c_1$ & AR parameter for leading speed & $\unit[0.010]{m/s}$\\
		$c_2$ & AR parameter for river depth & $\unit[0.262]{m}$\\
		$c_3$ & AR parameter for river cross-section & $\unit[4.992]{m^2}$\\
		$c_4$ & AR parameter for stream velocity & $\unit[0]{}$\\
		$\phi_1$ & AR parameter for leading speed & $\unit[0.994]{}$\\
		$\phi_2$ & AR parameter for river depth & $\unit[0.951]{}$\\
		$\phi_3$ & AR parameter for river cross-section & $\unit[0.997]{}$\\
		$\phi_4$ & AR parameter for stream velocity & $\unit[0.993]{}$\\
		$\sigma_1^2$ & AR variance parameter for leading speed & $\unit[0.034]{m^2/s^2}$\\
		$\sigma_2^2$ & AR variance parameter  for river depth & $\unit[0.381]{m^2}$\\
		$\sigma_3^2$ & AR variance parameter  for river cross-section & $\unit[598]{m^4}$\\	
		$\sigma_4^2$ & AR variance parameter  for stream velocity & $\unit[0.030]{m^2/s^2}$\\		
		\end{tabular}
\end{center}
\end{table}

\section{Validation}
\label{sec:validation}

To check if the trained model is safe, effective, and comfortable, we simulate different vessel-following scenarios. Instead of validating the model purely in scenarios based on the AR(1) processes used in training, we evaluated the generalization capabilities by also simulating scenarios that are not in the scope of the training data. In particular, we use real river dynamics from the Middle Rhine to validate our trained vessel-following model. A further aspect is to evaluate the string stability of the model by using a sequence of followers. Four different scenarios are described in the following.

\subsection{Scenario based on training AR(1) processes}
\begin{figure}[H]
    \centering
    \includegraphics[width=1\linewidth]{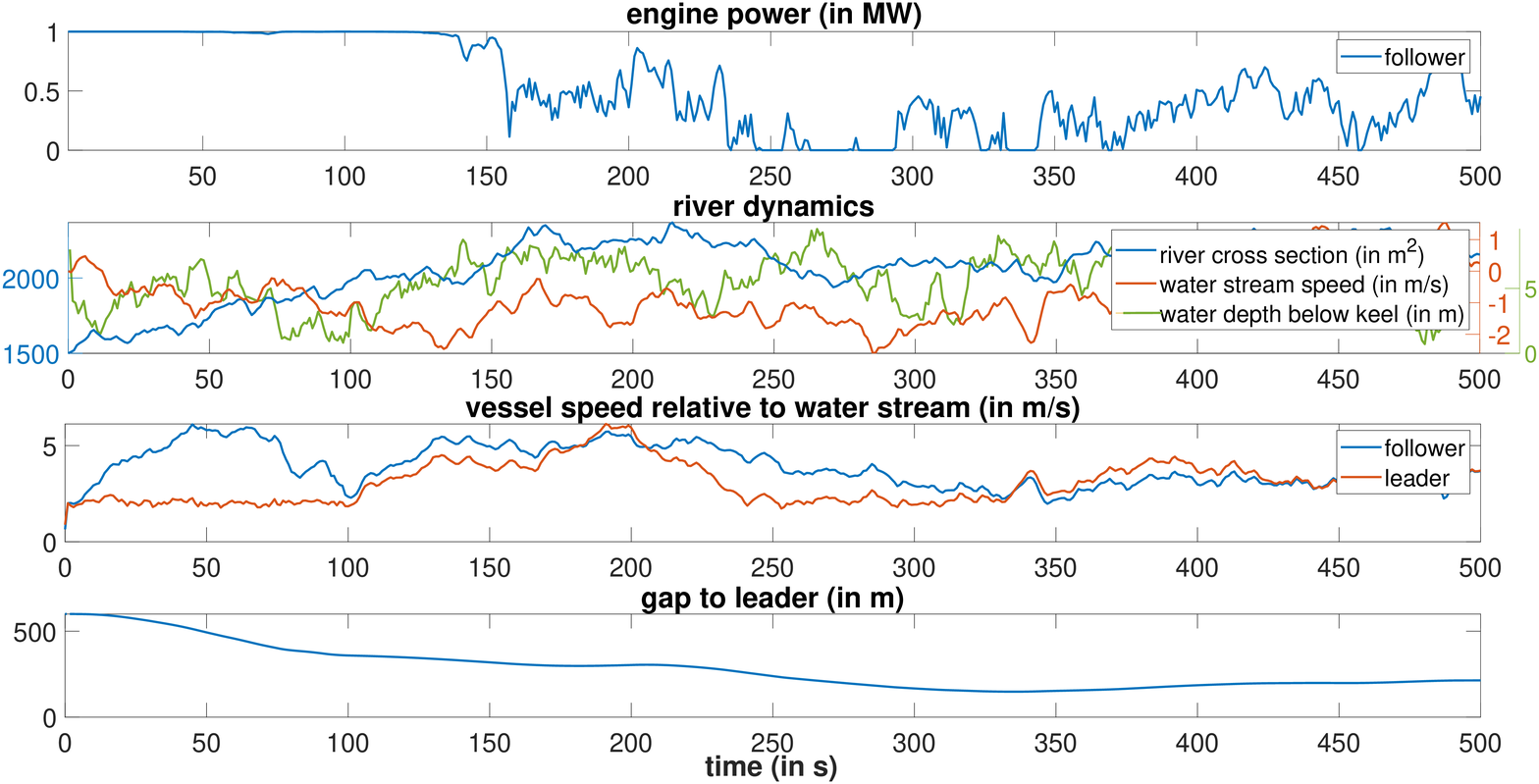}
    \caption{Validation scenario based on the AR(1) processes, that have been used in training.}
    \label{fig:scenario1}
\end{figure}

The first scenario is chosen similar to the training process in order to evaluate if the driving style is safe, effective, and comfortable. 
Figure \ref{fig:scenario1} shows the response of the trained model to a leader trajectory based on the AR(1) process we used in training. Furthermore, the river characteristics are also modeled by the training AR(1) processes. Both vessels start with a gap of \unit[600]{m} and the follower is moving with maximum power $P_{\rm max}$. When the gap falls below approximately \unit[400]{m}, the follower starts to slow down by decreasing the engine power and approaches the leader comfortably. During the whole scenario, the follower keeps a safe gap to the leader, although there are unrealistic high changes in river dynamics and leader speed. To compensate for these high changes, the follower's engine power shows high changing rates as well. Nevertheless, the gap to the leader never drops below \unit[150]{m}. It can be further observed that the spatial gap to the leader increases with the follower's speed, which is reflected in the reward structure (\ref{eq:timeheadwayReward}) that motivates the follower to keep a constant time gap. 

\subsection{Artificial scenario}
\begin{figure}[H]
    \centering
    \includegraphics[width=1\linewidth]{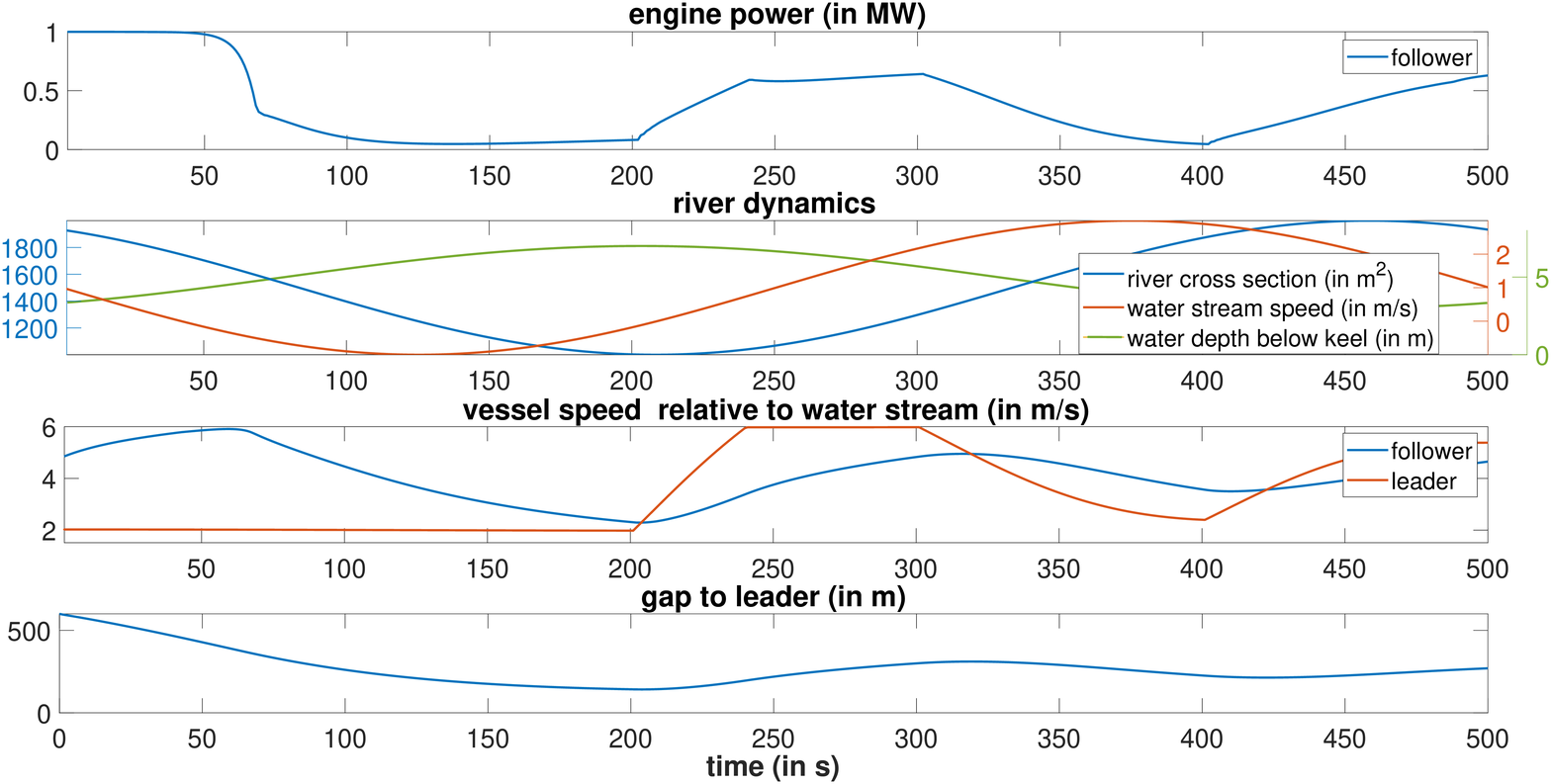}
    \caption{Artificial validation scenario based on sinus functions to model river dynamics and a realistic leading trajectory.}
    \label{fig:scenario2}
\end{figure}

Since the AR(1) processes used in the previous validation scenario are not suitable for modeling realistic environmental influences, we designed a more realistic scenario based on sinus functions to model river dynamics and by using a smoother leading trajectory (see Figure \ref{fig:scenario2}). As in the previous validation scenario, the initial gap between follower and leader is set to \unit[600]{m}. While in the beginning, the leader is moving with a minimum speed of \unit[2]{m/s} relative to water stream, the follower accelerates with maximum engine power $P_{\rm max}$. When the gap between both falls below \unit[400]{m}, the follower comfortably decreases its engine power and safely approaches the leading vessel with a final gap of approximately \unit[140]{m}. Thereafter, the leader accelerates and decelerates a few times while the follower reacts with comfortable changes in engine power and a safe gap to the leader. 

\subsection{Vessel-following on the Middle Rhine}
\begin{figure}[H]
    \centering
    \includegraphics[width=1\linewidth]{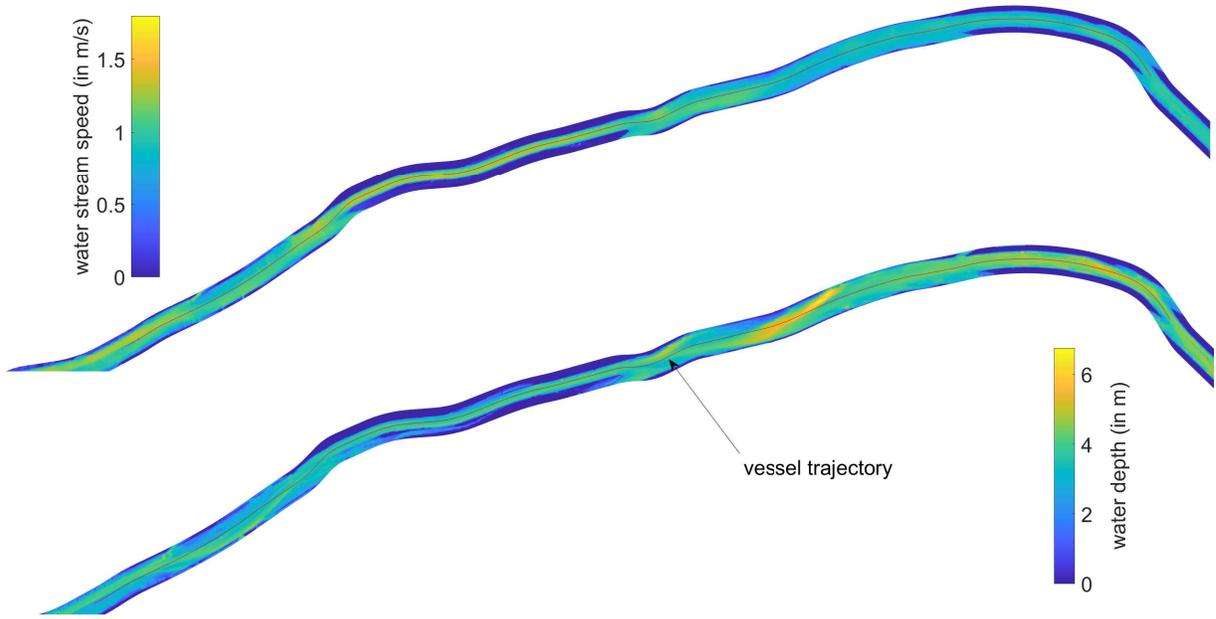}
    \caption{Chosen part of the Middle Rhine to validate the trained model. The upper illustration shows the water stream speed, the lower one the water depth. The vessel trajectory is marked in red.}
    \label{fig:river}
\end{figure}

To evaluate the agent's performance on a real river, we simulate a realistic vehicle following scenario using real river dynamics from the Rhine. In particular, we chose the part of the Middle Rhine that has been used to calibrate the reward function in Section \ref{sec:actionStateReward}. Figure \ref{fig:river} depicts the chosen river section with its geometry as well as water stream speed and water depth. The follower vessel trajectory is marked by red dots. 

\begin{figure}[H]
    \centering
    \includegraphics[width=1\linewidth]{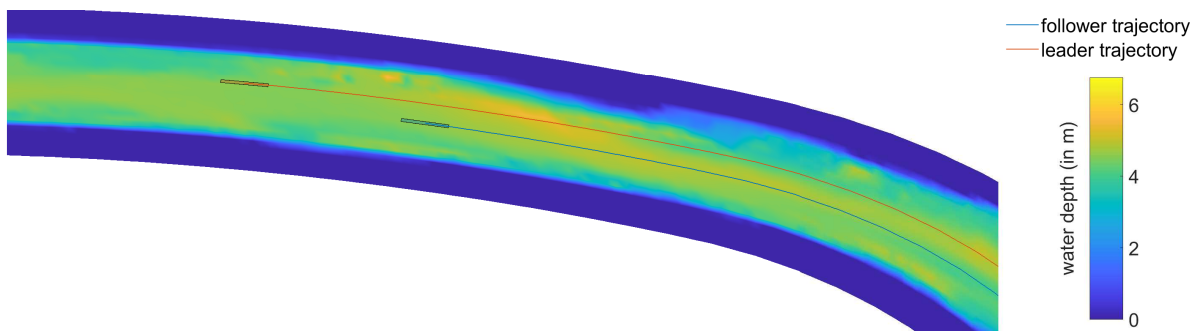}
    \caption{Lateral displacement of the follower with respect to the leader. This results in different river dynamics for both during simulation.}
    \label{fig:river_displacement}
\end{figure}

In contrast to the previous validation scenarios, we use the same vessel dynamics (cf. Section \ref{sec:vesselDynamics}) for the leader that we used for the follower. Since, in reality, engine powers are aimed to be kept constant, we set a constant engine power of $P = \unit[0.5]{MW}$ for the leading vessel in this validation scenario. Furthermore, we set the follower with a constant lateral displacement with respect to the leader, depicted in Figure \ref{fig:river_displacement}. This results in a scenario that is, on the one hand, more realistic since vessels usually do not travel directly behind each other and, on the other hand, more challenging since river dynamics are different for both vessels during simulation.

\begin{figure}[H]
    \centering
    \includegraphics[width=1\linewidth]{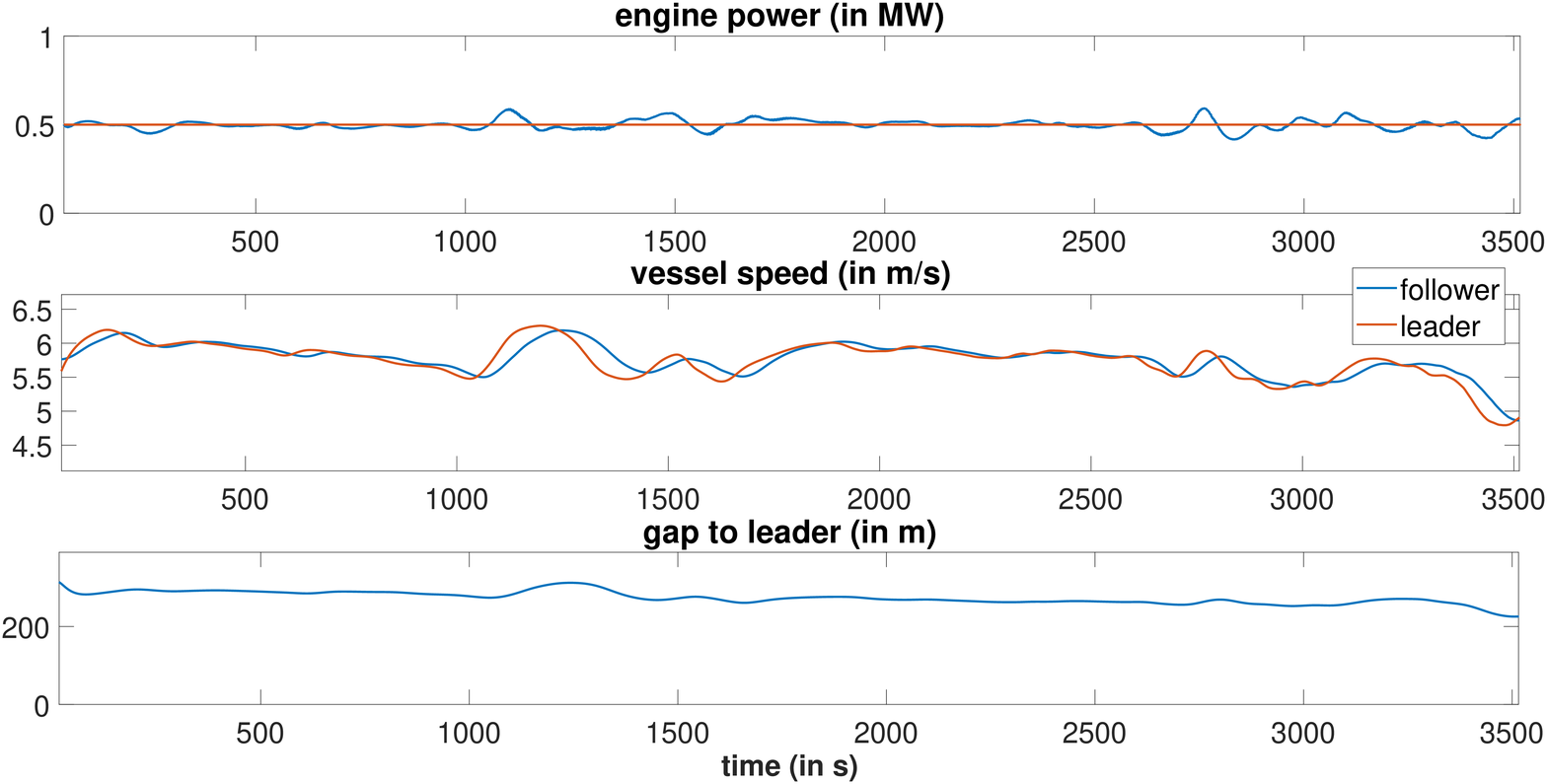}
    \caption{Response of the trained model to a leading vessel, that is travelling with constant engine power on the Middle Rhine.}
    \label{fig:scenario3}
\end{figure}

Figure \ref{fig:scenario3} shows the response of the follower to the leader that is traveling downstream with constant engine power. Since follower and leader are experiencing different environmental influences, the follower has to adjust its engine power. But again, the changing rates in engine power are low, resulting in a comfortable driving style. The following vessel is able to follow the speed of the leader quite well and is, therefore, able to keep a safe gap to the leader that never drops below \unit[200]{m}.

\subsection{Sequence of followers on the Middle Rhine}
\begin{figure}[H]
    \centering
    \includegraphics[width=1\linewidth]{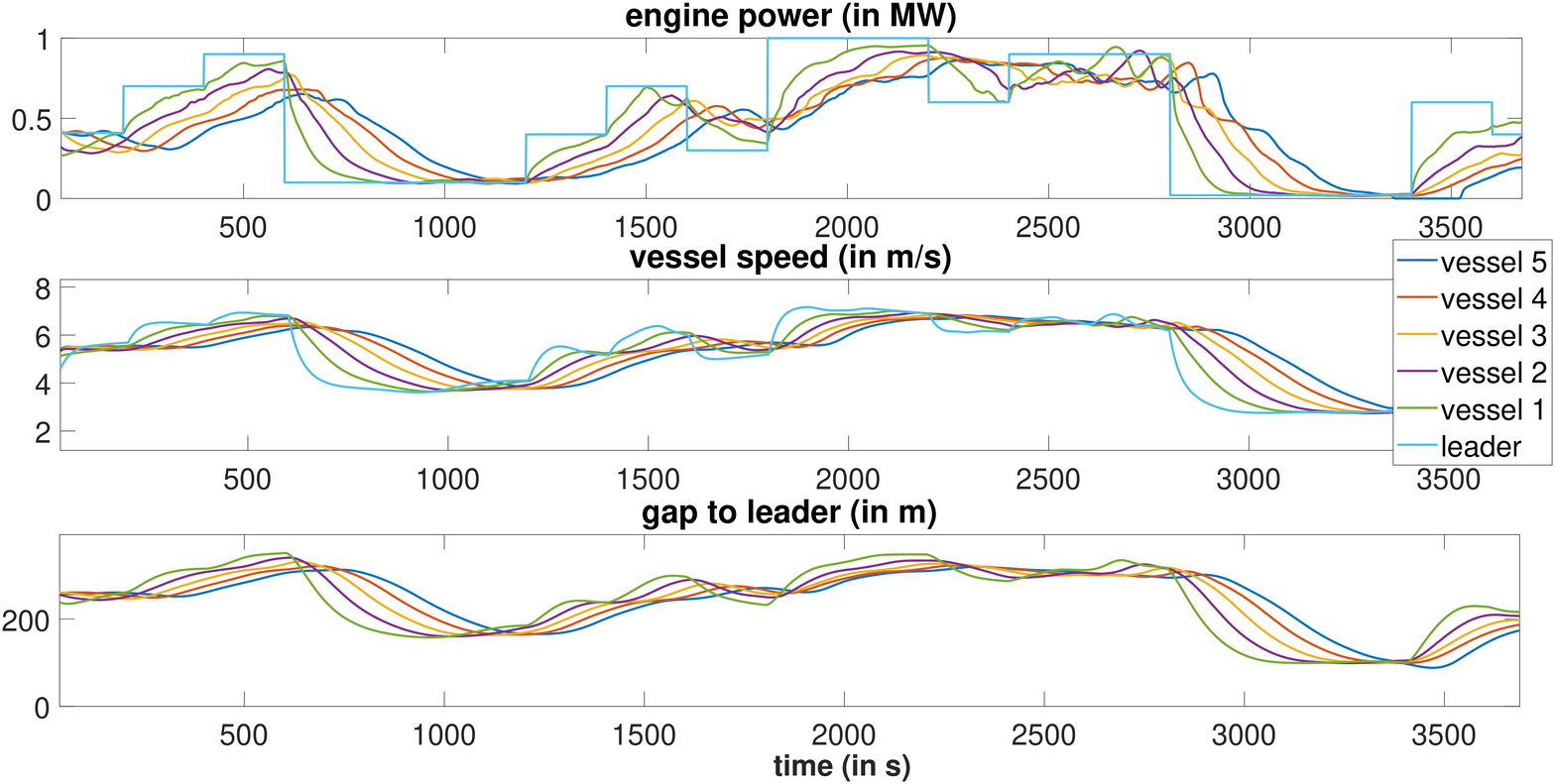}
    \caption{Reaction of a sequence of five followers, each controlled by the trained model, on the Middle Rhine.}
    \label{fig:scenario4}
\end{figure}

This vessel following scenario is designed to test two aspects: First, we aim to evaluate how the trained model reacts on a leading vessel that shows high jumps in engine power, resulting in high accelerations and safety-critical decelerations. At this point, it is again worth mentioning that we do not consider negative engine power. Therefore, the most safety-critical situation is defined by a sudden and high decrease in the leading vessel's engine power. The second aspect to evaluate is string stability. This is realized by using a sequence of five followers traveling behind a leading vessel with zero lateral offset. All five followers use the trained model to control their engine power, and again we use a part of the Middle Rhine. 
Figure \ref{fig:scenario4} shows the reaction of the five followers to the leading trajectory that is based on jumps in engine power. All vessels are able to keep a safe gap to the respective leader and further show comfortable changing rates in engine power. In the most critical situation, when the leading vehicle suddenly reduces its engine power to almost zero at $t \approx \unit[2800]{s}$, the followers react quickly but still with a comfortable decrease in engine power, enabling them to keep safe gaps to their respective leaders. Furthermore, string stability can be observed in a way that no oscillations occur. On the contrary, the sequence of followers is flattening the speed profile and thus is able to dampen oscillations and increase comfort.

\subsection{Comparison with real leader-follower pair}
\begin{figure}[H]
    \centering
    \includegraphics[width=1\linewidth]{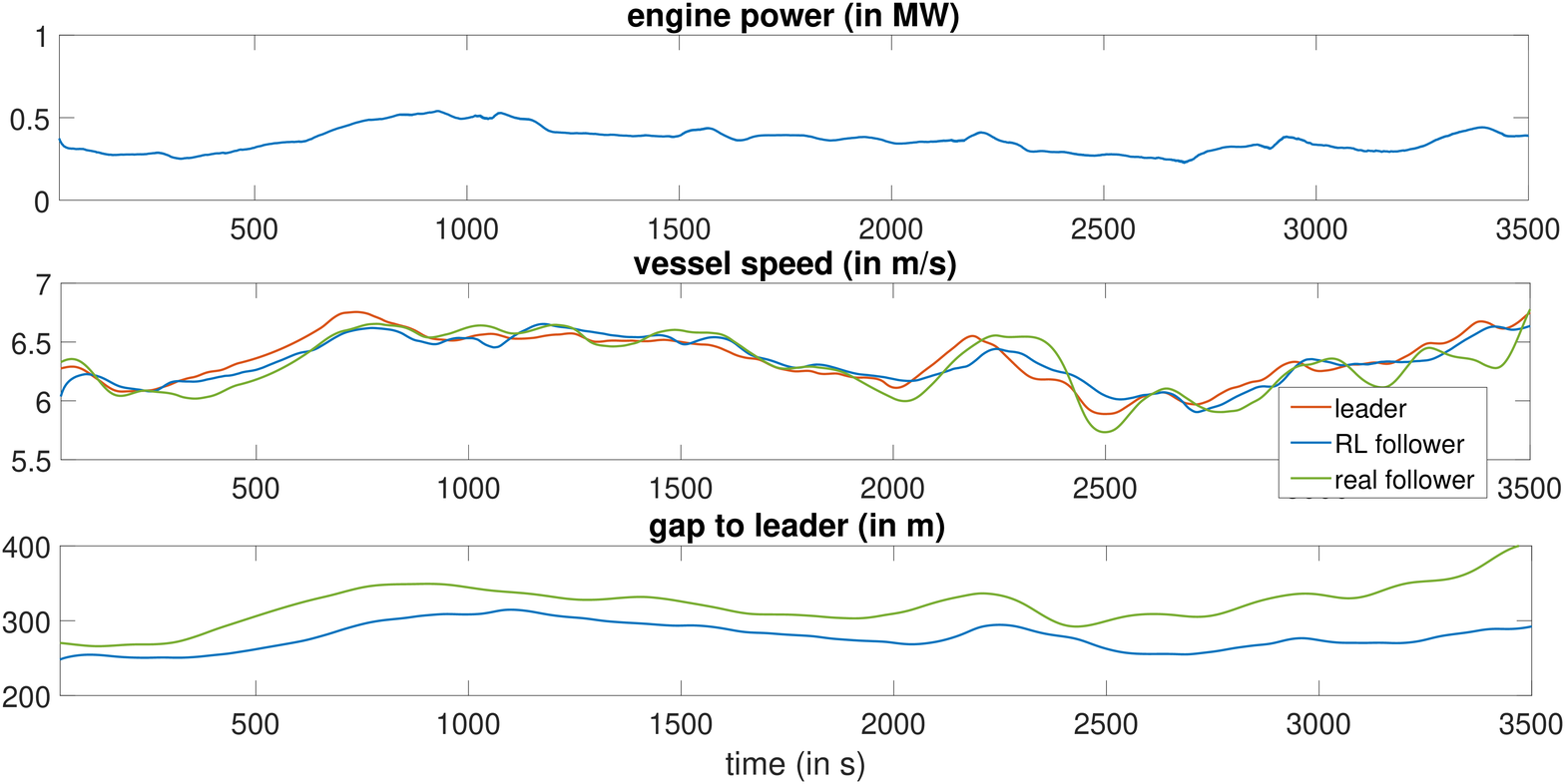}
    \caption{Comparison of the trained agent with a follower-leader pair from the AIS data set.}
    \label{fig:scenario5}
\end{figure}

In a last experiment, we compare the behavior of the trained model with real vessel-following behavior. Figure \ref{fig:scenario5} depicts a scenario where we took a real follower-leader pair from the AIS data set and let our trained agent follow the exact same leading trajectory. Comparing real and RL follower trajectories, we see that they both show roughly the same behavior regarding their speed curve, remarking that the RL follower shows less variance from around $t = \unit[2000]{s}$. By comparing the resulting gap to the leader, we can further observe that the real follower chooses overall higher gaps than the RL follower. This can be interpreted as a more defensive driving style that we can also see in road traffic. However, it is to mention that such a different behavior can easily be achieved by shifting the lognormal distribution in the safety factor of the reward function (\ref{eq:timeheadwayReward}) towards larger bow-stern time gaps. 

\section{Conclusion}
\label{sec:Conclusion}

This study presents an RL-based vessel-following model taking into account realistic vessel dynamics, including environmental influences, such as stream velocity and river profile. For the formulation of a suitable reward, we extracted natural vessel behavior from AIS data from a part of the Middle Rhine. Using these insights, we defined a reward function that reflects realistic vessel-following behavior as well as safe and comfortable driving. In order to guarantee collision-free navigation on different types of rivers, we developed a general training environment that uses AR(1)-processes to model the leading vessel trajectory as well as changing river dynamics. 

We evaluated the performance of the trained model in different scenarios ranging from artificial and unrealistic waterways to realistic vessel-following on the Middle Rhine. To validate the generalization capabilities, all of these scenarios have never been seen in training. Although some scenarios were designed to bring the model to its limits, the model proved to be accident-free while maintaining a comfortable driving style in all situations. Furthermore, the trained model was able to effectively dampen traffic oscillations in a sequence of trained followers, even if the leader showed extreme acceleration values. Since this aspect was neither trained nor included in the agent's specification, this is another proof of the generalization abilities. 

In conclusion, we showed that RL can not only handle simple point-mass dynamics in vehicle-following but is able to perform well on tasks where vehicle dynamics are complex and under the influence of strong external disturbances. Based on these insights, we plan to extend this study by using a two-dimensional vessel model in the future. Challenges that come along with this task would be the consideration of a two-dimensional action space, including engine power and rudder angle, as well as additional requirements regarding lateral navigation.

\subsubsection*{Acknowledgements}
This work was funded by BAW - Bundesanstalt für Wasserbau (Mikrosimulation des Schiffsverkehrs auf dem Niederrhein).


\appendix


\bibliographystyle{elsarticle-harv}
\bibliography{ShipFollowing}


%
%
%
\end{document}